# Design of a Low Voltage Class-AB CMOS Super Buffer Amplifier with Sub Threshold and Leakage Control

Rakesh Gupta

Assistant Professor, Electrical and Electronic Department, Uttar Pradesh Technical University, Lucknow
Uttar Pradesh, India

**Abstract--**
**This paper describes a CMOS analogy voltage supper buffer designed to have extremely low static current Consumption as well as high current drive capability. A new technique is used to reduce the leakage power of class-AB CMOS buffer circuits without affecting dynamic power dissipation. The name of applied technique is TRANSISTOR GATING TECHNIQUE, which gives the high speed buffer with the reduced low power dissipation (1.105%), low leakage and reduced area (3.08%) also. The proposed buffer is simulated at 45nm CMOS technology and the circuit is operated at 3.3V supply[11]. Consumption is comparable to the switching component. Reports indicate that 40% or even higher percentage of the total power consumption is due to the leakage of transistors. This percentage will increase with technology scaling unless effective techniques are introduced to bring leakage under control. This article focuses on circuit optimization and Design automation techniques to accomplish this goal [9].**

## I. INTRODUCTION

The input offset voltage of a practical operational amplifier consists of a random and a systematic part. The random errors are caused by random device mismatches. Systematic errors can be considered as errors in the design. The most commonly used as amplifier is common emitter (Bipolar junction transistor) or common source (MOSFET) that magnifies and invert the input signal. [1] Behzad Rajavi says that we know that MOSFET plays an important role in the reduced size of the chip, by reducing the gate oxide thickness of MOSFET *i.e.*, (Tox). But reducing of the Tox gives the reduced tolerance of the MOSFET devices for the higher voltage levels at the gate of MOSFET. It means that to reduce the maximum supply voltages Vdd it gives the helpful purpose. Due to the reduced supply voltages analogue designer have to face some common problems like input common mode range, output swing, and linearity of the device. In the resulting form to implement the desired analogue device we apply the CMOS technology with low voltage and low power techniques. Voltage supper buffers are essential building blocks in analog and mixed-signal circuits and processing systems, especially for applications where the weak signal needs to be delivered to a large capacitive load without being distorted To achieve higher density and performance and lower power consumption, CMOS devices have been scaled for more than 30 years. Transistor delay times have decreased by more than 30% per technology generation resulting in doubling of microprocessor performance every two years. Supply voltage ( VDD) has been scaled down in order to keep the power consumption under control. Hence, the transistor threshold voltage (Vth) has to be commensurately scaled to maintain a high drive current and achieve the performance improvement. However, the threshold voltage scaling results in the substantial increase of the sub threshold leakage current.

## II. TRANSISTOR LEAKAGE MECHANISMS

Six short channel leakage mechanisms are illustrated in Fig. 2 $I_1$ is the reverse bias *pn* junction leakage; $I_2$ is the subthreshold leakage; $I_3$ is the oxide tunnelling current; $I_4$ is the gate current due to hot carrier injection; $I_5$ is the Gate Induced Drain Leakage (GIDL); and $I_6$ is the channel punch through current. Currents $I_2$, $I_5$, $I_6$ are off-state leakage mechanisms while $I_1$ and $I_3$ occur in both ON and OFF states. $I_4$ can occur in the off-state, but more typically occurs during the transistor bias states in transition.

*A. pn Junction Reverse Bias Current*

Drain and source to well junctions are typically reverse biased causing *pn* junction leakage current. A reverse bias *pn* junction leakage ($I_1$) has two main components: One is minority carrier diffusion/drift near the edge of the depletion region and the other is due to electron hole pair generation in the depletion region of the reverse biased junction. For an MOS transistor, additional leakage can occur between the drain and well junction from gated diode device action (overlap of the gate to the drain-well *pn* junctions) or carrier generation in drain to well depletion regions with influence of the gate on these current components. *pn* junction reverse bias leakage (IREV) is a function of junction area and doping concentration. If both *n*- and *p*-regions are heavily doped (this is the case for advanced MOSFETs using heavily doped shallow junctions and halo(doping), Band-To-Band Tunnelling (BTBT) dominates the pn junction leakage [2].

*B. Subthreshold Current*

Circuit speed improves with increasing Ion, therefore it would be desirable to use a small Vt. Can we set Vt at an arbitrarily small value, say 10mV? The answer is no.At Vgs<Vt, an N-channel MOSFET is in the off-state. However, an undesirable leakage current can flow between the drain and the source. The MOSFET current observed at Vgs<Vt is called the subthreshold current. This is the main contributor to the MOSFET off-state current, Ioff. Ioff is the Id measured at Vgs=0 and Vds=Vdd. It is important to keep Ioff very small in order to minimize the static power that a circuit consumes even when it is in the standby mode. For example, if Ioff is a modest 100nA per transistor, a cell-phone chip containing one hundred million transistors would consume so much standby current (10A) that the battery would be drained in minutes without receiving or transmitting any calls. A desk-top PC chip may be able to tolerate this static power but not much more before facing expensive problems with cooling the chip and the system. Fig. 1 shows a typical subthreshold current plot. It is almost always plotted in a semi log Ids versus Vgs graph. When Vgs is below Vt, Ids is an exponential function of Vgs.

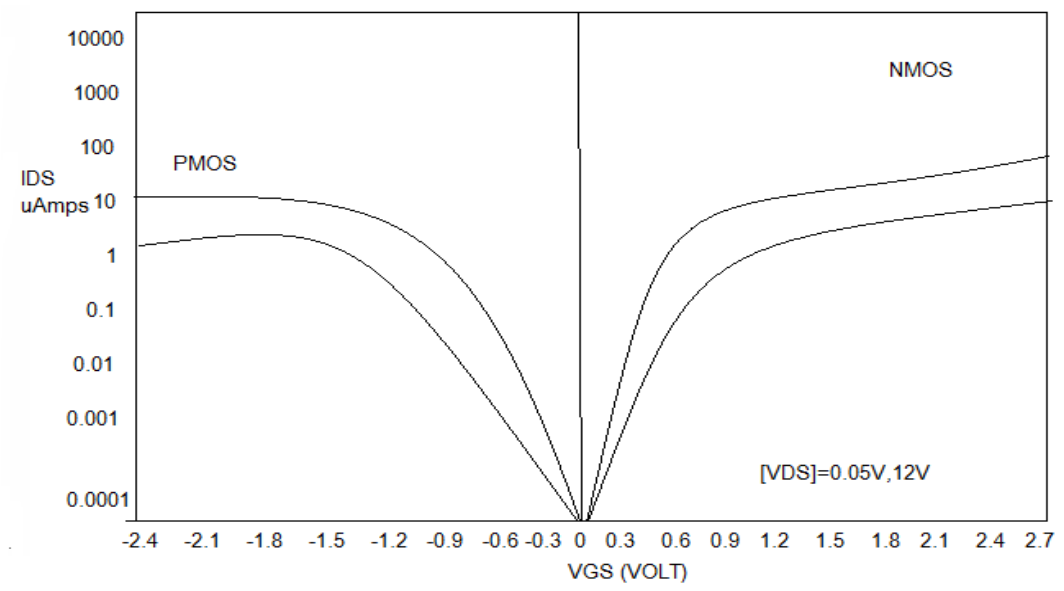

Fig. 1 The current that flows at Vgs<Vt is called the subthreshold current. Vt ~0.2V.The lower/upper curves are for Vds=50mV/2.7V

*C. Sub threshold Leakage*

The *sub threshold leakage* is the drain-source current of a transistor operating in the weak Inversion region. Unlike the strong inversion region in which the drift current dominates, the sub threshold conduction is due to the diffusion current of the minority carriers in the Channel for a MOS device.1 For instance, in the case of an inverter with a low input voltage, the NMOS is turned OFF and the output voltage is high[5]. In this case, although VGS is 0V, there is still a current flowing in the channel of the OFF NMOS transistor due to the VDD potential of the VDS. The magnitude of the sub threshold current is a function of the temperature, supply voltage, device size, and the process parameters out of which the threshold voltage (VT) plays a dominant role.

$$I_{sh} = \frac{Kn}{2}$$

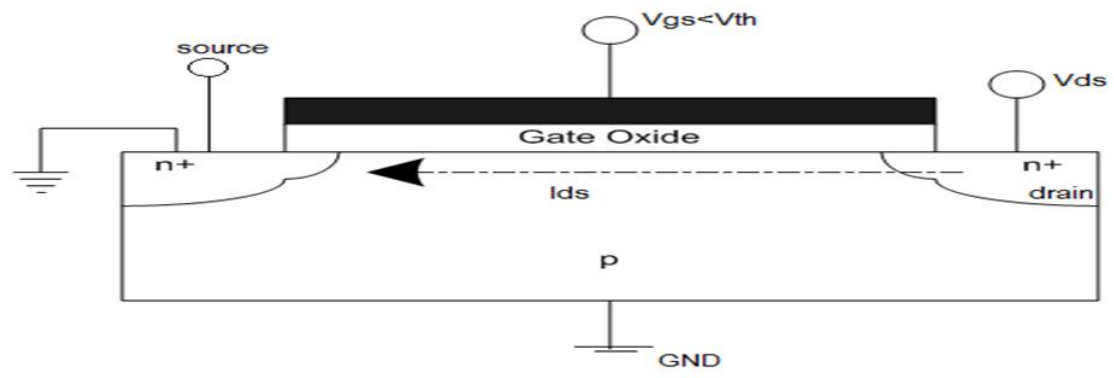

Fig. 2 Leakage current mechanisms of deep submicron Transistors.

Subthreshold or weak inversion conduction current between source and drain in an MOS transistor occurs when gate voltage is below Vth [2]. Weak inversion typically dominates modern device off-state leakage due to the low Vth. The weak inversion current can be expressed based on the following equation [2][6]:

$$I_{ds} = \mu_o C_{gs}\, W/L (m-1)(V_T)^2 \text{ x } e^{\frac{Vg-Vth}{Mvr}} (1 - e^{-\frac{Vds}{Vy}})$$

$$M = 1 + \frac{3.tox}{Wdm}$$

Where $V_T$ is the temperature voltage derived from

$$\mathbf{V_T} = \frac{k.T}{q}$$

K= Boltzmann constant
T= absolute temperature and q = electron charge
*Vth* = threshold voltage, *Cgs* is the gate oxide capacitance; µ*0* is the zero bias mobility
M = sub threshold swing coefficient (also called body effect coefficient)
*Wdm* = maximum depletion layer width
*tox* = gate oxide thickness

In long channel devices, the subthreshold current is independent of the drain voltage for *VDS* larger than few *vT*. On the other hand, the dependence on the gate voltage and threshold voltage is exponential. In long-channel devices, the source and drain are separated far enough that their depletion regions have no effect on the potential or field pattern in most part of the device[7]. Hence, for such devices, the threshold voltage is virtually independent of the channel length and drain bias. In a short channel device, however, the source and drain depletion width in the vertical direction and the source-drain potential have a strong effect on the band bending over a significant portion of the device [12]. Therefore, the threshold voltage and consequently the subthreshold current of short channel devices vary with the drain bias.

*D. Subthreshold Leakage Current Control Method*

At Vgs below Vt, the inversion electron concentration (ns) is small but nonetheless can allow a small leakage current to flow between the source and the drain. In Fig. 7-2(a), a large Vgs would pull the Ec at the surface closer to Ef, causing ns and Ids to rise. From the equivalent circuit in Fig. 2.4, one can observe that

$$\frac{d\emptyset s}{dvgs} = \frac{Coxe}{Coxe+Cdep} = \frac{1}{\eta} \quad \ldots\ldots (1)$$

$$\eta = 1 + \frac{Cdep}{Coxe} \quad \ldots\ldots (2)$$

Integrating Eq. (1) yields

$$\emptyset s = \text{constant} + V/h \quad \ldots\ldots (3)$$

$I_{ds}$ is proportional to ηs, therefore

$$I_{ds} \propto \eta s \propto e^{q\emptyset s/KT} \quad \ldots\ldots (4)$$

The practical definition of Vt in experimental studies is the Vgs at which $I_d$s=100nA ×W/L.Eq. (4) may be rewritten as

$$Ids_{(nA)} = 100.\frac{W}{L}.e^{q(Vgs-Vt)/\eta KT} \quad \ldots\ldots (5)$$

Clearly, Eq. (5) agrees with the definition of Vt and Eq. (4). Recall that the function exp(qVgs/kT) changes by 10 for every 60 mV change in Vgs, therefore exp(qVgs/ hkT) changes by 10 for every h×60mV. For example, if h=1.5, Eq. (5) states that Ids drops by 10 times for every 90mV of decrease in Vgs below Vt. h×60mV is called the subthreshold swing and represented by the symbol, S.

$$S = \eta.60mv.\frac{T}{300} \quad \ldots\ldots (6)$$

$$Ids_{(nA)} = 100.\frac{W}{L}.e^{q(Vgs-Vt)/\eta KT} \quad \ldots\ldots (7)$$

$$Ioff_{(nA)} = 100.\frac{W}{L}.e^{-qVt)/\eta KT} \quad \ldots\ldots (8)$$

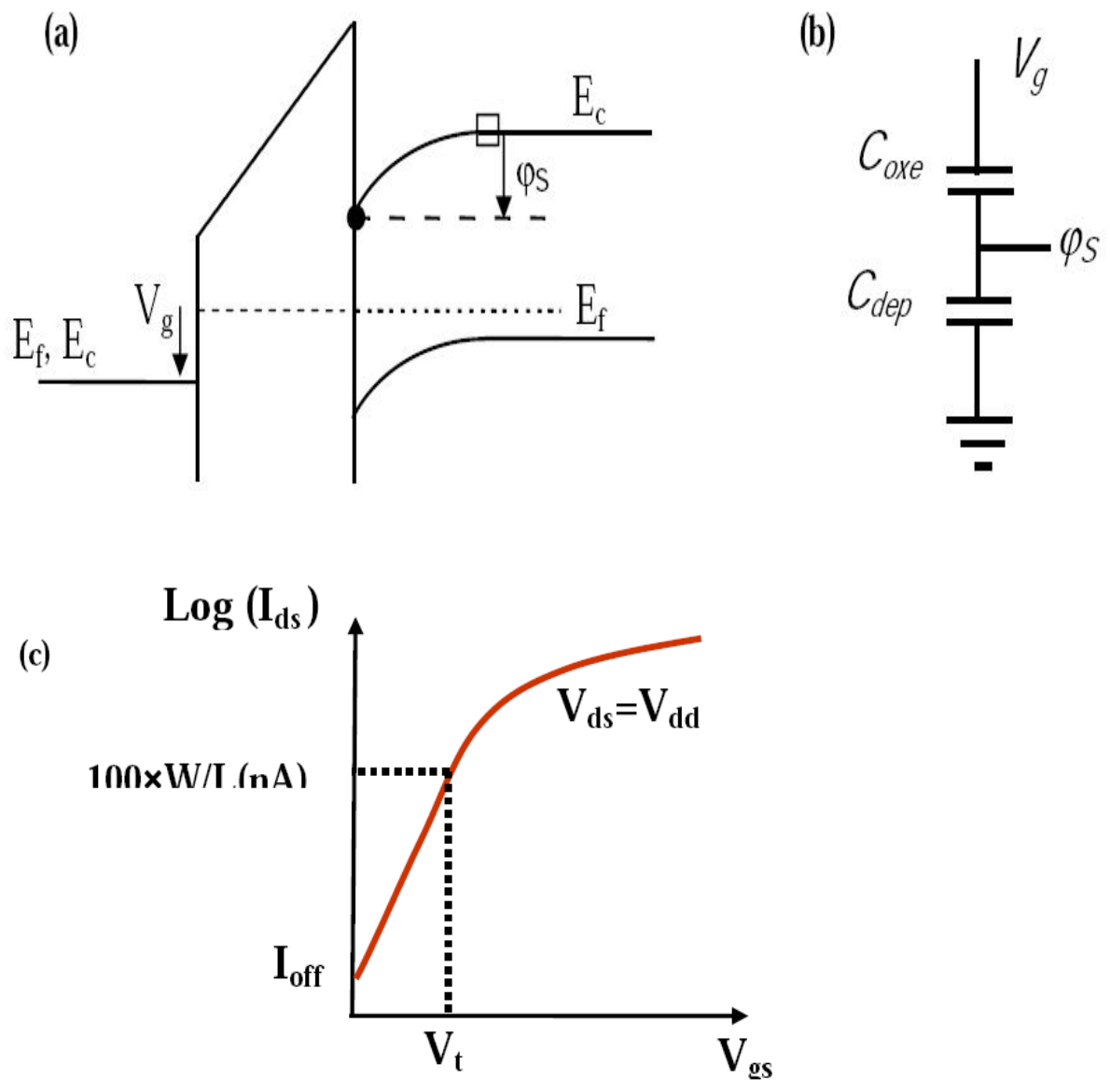

Fig.3 (a) When Vg is increased, Ec at the surface is pulled closer to Ef, causing ns and Ids to rise; (b) equivalent capacitance network; (c) Subthreshold IV with Vt and Ioff.

For given W and L, there are two ways to minimize Ioff illustrated in Fig. 3(c). The first is to choose a large Vt. This is not desirable because a large Vt reduces Ion and therefore increases the gate delays [3]. The preferable way is to reduce the subthreshold swing. S can be reduced by reducing h. That can be done by increasing Coxe (see Eq. 2), i.e. using a thinner Tox, and by decreasing Cdep, i.e. increasing Wdep.2 An additional way to reduce S, and therefore to reduce Ioff, is to operate the transistors at a lower temperature. This last approach is valid in principal but rarely used because cooling adds considerable cost [4].

## III. DESIGN OF SUPPER BUFFER AMPLIFIER

These devices provide the designer with high performance operation at low supply voltages and selectable quiescent currents, and are an ideal design tool when ultra-low input current and low power dissipation are desired. Design consisted PMOS (PQ1,PQ2,PQ3………PQ9) and NMOS(NQ1,NQ2………NQ11) CMOS transistor which control total function of buffer amplifier . Three steps involved to proper show accuracy of Circuit input stage is first and more accurate block design in this circuit , CMOS PQ1 and NQ1 work as invertor in circuit and generated low error leakage in the total function of circuit.

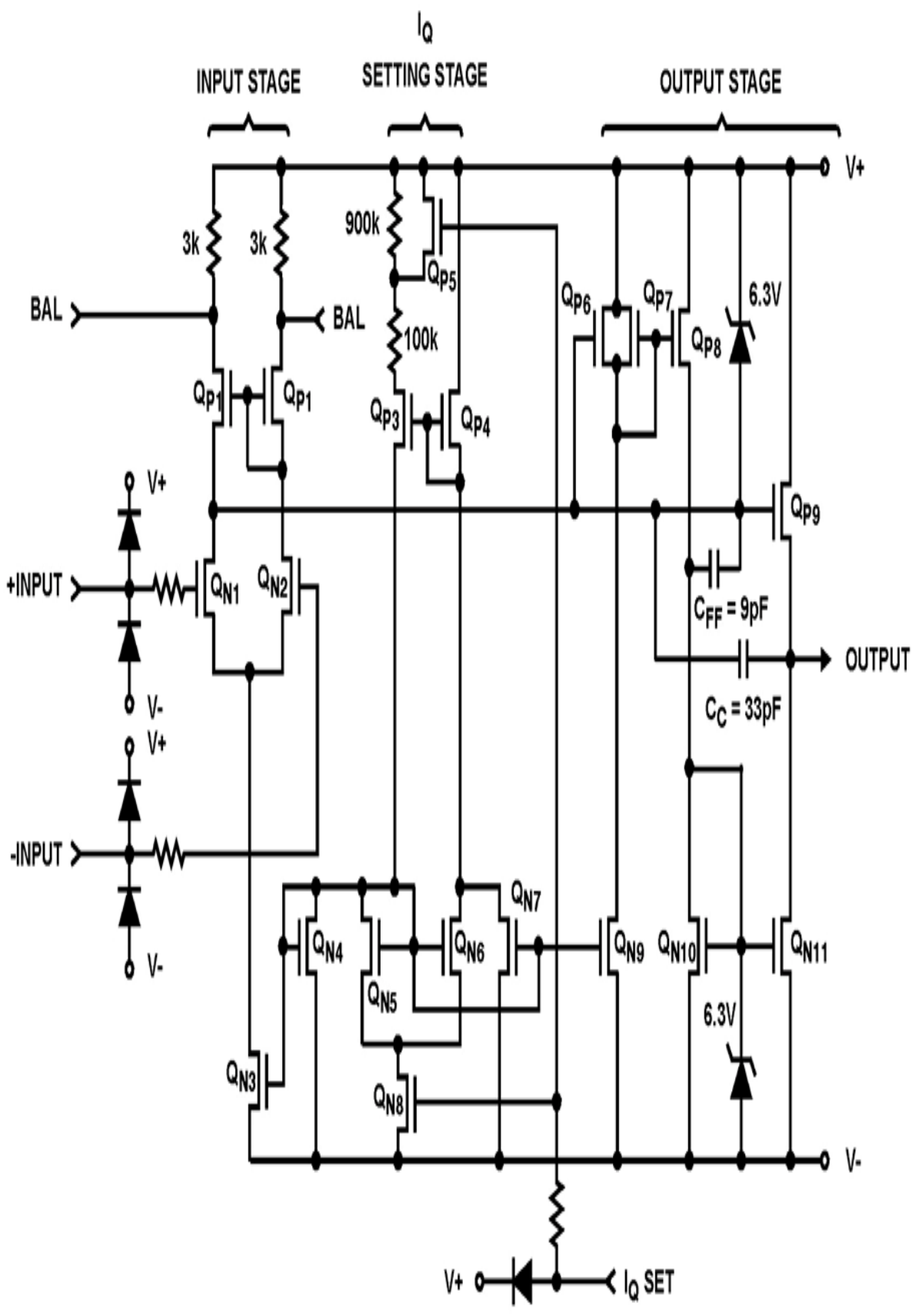

Fig. 4 Super Buffer Amplifier Circuit model

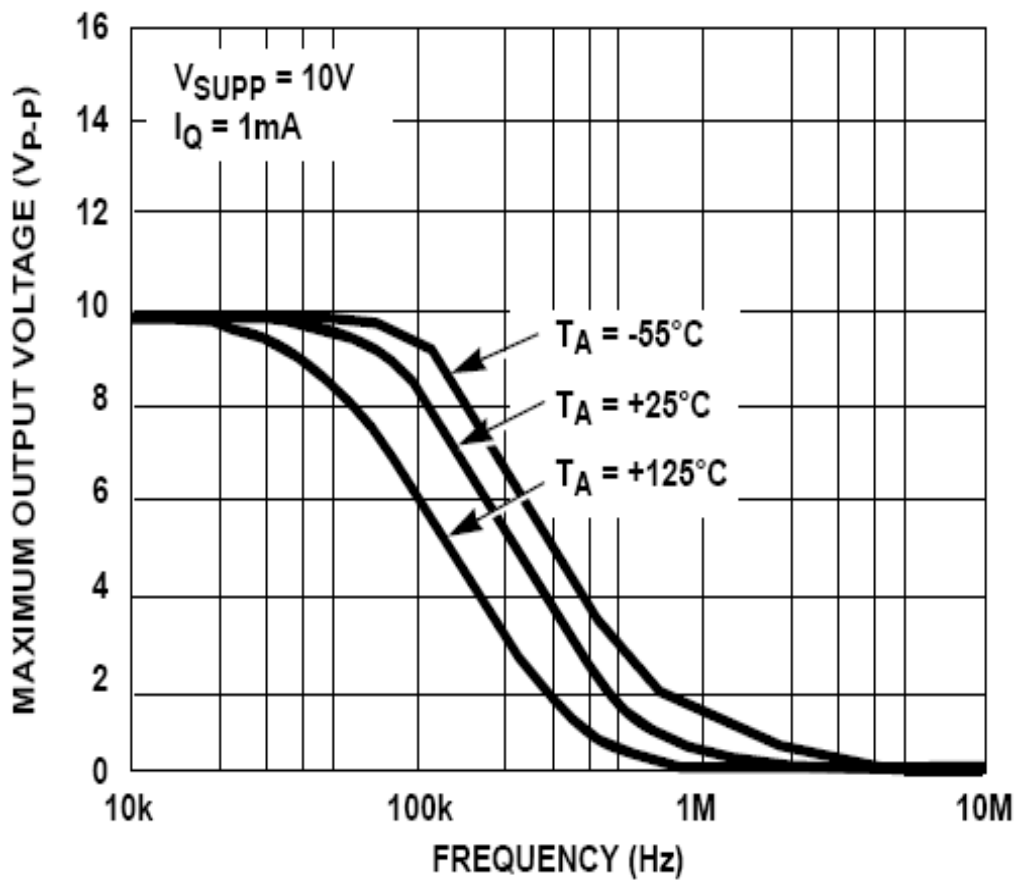

Fig. 5 wave model between $V_{out_max}$ vs Frequency

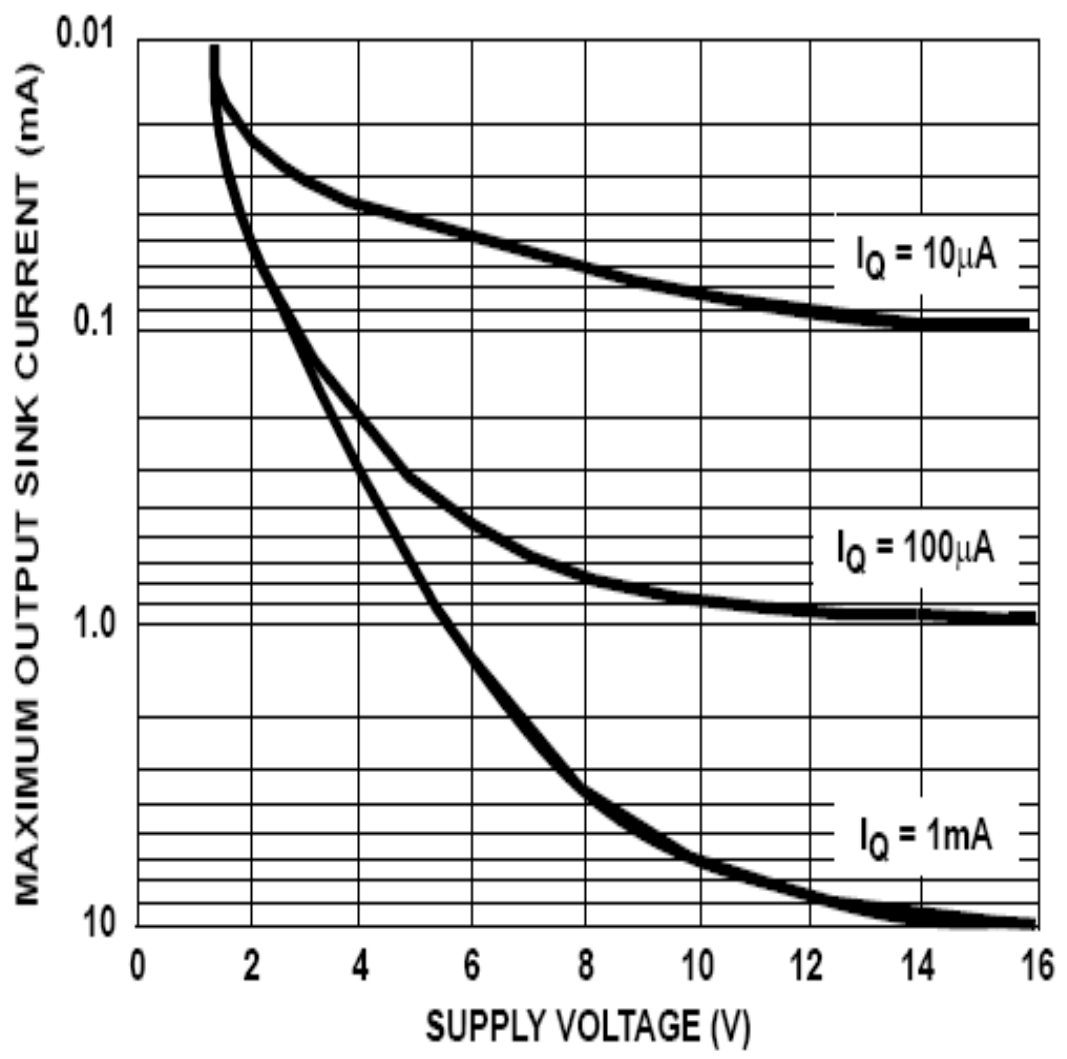

Fig. 6 Output sink vs supply voltage

## IV. SIMULATION RESULTS

Amplifier using 45nm Buffer CMOS technology we designed a new buffer as shown in Fig. 4. which is simulated at 3V supply voltage by help of the cadence tool. Fig. 4 consists the transistors that all have the same sizing. Bias current IB is fixed at 10µA in buffer circuit.

| PARAMETERS | |
|---|---|
| Slew rate | 1.6V/µs |
| Bandwidth | 1.4MHZ |
| Noise | 100nV/HZ |
| Iout | 40mA |
| Rail to Rail Output | Low |
| Transistor count | 20 |

## V. CONCLUSION

As technology continues to scale, subthreshold leakage currents will increase exponentially. Estimates on microprocessors show that subthreshold leakage currents can easily consume upwards of 30% of the total power budget in an aggressive technology. In the past, circuit techniques and architectures ignored the effects of these currents because they were insignificant compared to dynamic currents since threshold voltages were so high. With the continuous scaling of CMOS devices, leakage current is becoming a major contributor to the total power consumption. In current deep submicron devices with low threshold voltages, subthreshold and gate leakage have become dominant sources of leakage and are expected to increase with the technology scaling. To manage the increasing leakage in deep-submicron CMOS circuits, solutions for leakage reduction have to be sought both at the process technology and circuit levels. The settling time of proposed circuit is also reduced to the range of nanoseconds. This technique is also capable to enhance the slew rate, the achieved slew rate is 90(v/µs).The designed buffers is applicable in systems requiring the efficient operation with very low quiescent power consumption.